\documentclass[letterpaper]{article} % DO NOT CHANGE THIS
\usepackage{aaai2026}  % DO NOT CHANGE THIS
\usepackage{times}  % DO NOT CHANGE THIS
\usepackage{helvet}  % DO NOT CHANGE THIS
\usepackage{courier}  % DO NOT CHANGE THIS
\usepackage[hyphens]{url}  % DO NOT CHANGE THIS
\usepackage{graphicx} % DO NOT CHANGE THIS
\usepackage{natbib}  % DO NOT CHANGE THIS AND DO NOT ADD ANY OPTIONS TO IT
\usepackage{caption} % DO NOT CHANGE THIS AND DO NOT ADD ANY OPTIONS TO IT
\usepackage{algorithm}
\usepackage{algorithmic}

\usepackage{newfloat}
\usepackage{listings}
\DeclareCaptionStyle{ruled}{labelfont=normalfont,labelsep=colon,strut=off} % DO NOT CHANGE THIS
\floatstyle{ruled}
\newfloat{listing}{tb}{lst}{}
\floatname{listing}{Listing}
\usepackage{inconsolata}

\usepackage{subfiles}
\usepackage{array}
\usepackage{threeparttable}
\usepackage{booktabs}
\usepackage{siunitx} % line up decimal points
\usepackage{bm}

\usepackage{tabularx}
\newcolumntype{k}{X}
\newcolumntype{s}{>{\hsize=.35\hsize}X}
\newcolumntype{y}{>{\hsize=.25\hsize}X}
\usepackage{makecell}

\usepackage{amssymb}% http://ctan.org/pkg/amssymb
\usepackage{amsmath} 

\usepackage{tikz}
\usetikzlibrary{arrows.meta, positioning, shapes.geometric}

\usepackage{color}
\definecolor{myorange}{RGB}{240, 96, 0}

\title{Self-reflection in Automated Qualitative Coding: Improving Text Annotation through Secondary LLM Critique}

\author{
    Zackary Okun Dunivin\textsuperscript{\rm 1}, 
    Mobina Noori\textsuperscript{\rm 2}, 
    Seth Frey\textsuperscript{\rm 2}, 
    Curtis Atkinson\textsuperscript{\rm 3}
}
\affiliations{
    \textsuperscript{\rm 1}University of Stuttgart \quad \textsuperscript{\rm 2}University of California Davis \quad \textsuperscript{\rm 3}University of Washington \\
    \textit{zackary.dunivin@sowi.uni-stuttgart.de}\\
    \vspace{1em}
    {\large \today}

}

\nocopyright
\begin{document}

\maketitle

\begin{abstract}
Large language models (LLMs) allow for sophisticated qualitative coding of large datasets, but zero-/few-shot classifiers can produce an intolerable number of errors, even with careful, validated prompting. We present a simple, generalizable two-stage workflow: an LLM applies a human-designed, LLM-adapted codebook; a secondary LLM critic performs self-reflection on each positive label by re-reading the source text alongside the first model’s rationale and issuing a final decision. We evaluate this approach on six qualitative codes over 3,000 high-content emails from Apache Software Foundation project evaluation discussions. Our human-derived audit of 360 positive annotations (60 passages by six codes) found that the first-line LLM had a false-positive rate of 8\%--54\%, despite F1 scores of 0.74--1.00 in testing. 
Subsequent recoding of all stage-one annotations via a second self-reflection stage improved F1 by 0.04--0.25, bringing two especially poor performing codes up to 0.69 and 0.79 from 0.52 and 0.55 respectively. Our manual evaluation identified two recurrent error classes: misinterpretation (violations of code definitions) and meta-discussion (debate about a project evaluation criterion mistaken for its use as a decision justification). Code-specific critic clauses addressing observed failure modes were especially effective with testing and refinement, replicating the codebook-adaption process for LLM interpretation in stage-one. We explain how favoring recall in first-line LLM annotation combined with secondary critique delivers precision-first, compute-light control. With human guidance and validation, self-reflection slots into existing LLM-assisted annotation pipelines to reduce noise and potentially salvage unusable classifiers.
\end{abstract}

\section{Introduction}

Large language models (LLMs) have rapidly moved from speculative tools to practical partners in qualitative coding and text annotation. Social scientists now have abundant worked examples showing that frontier and near-frontier models can match or approach expert coders on conceptually rich tasks when prompts and codebooks are carefully designed and validated. Dunivin \citeyearpar{dunivin2025scaling} demonstrates how adapting human codebooks for machine comprehension---tightening definitions, adding boundary cases, and using structured, rationale-eliciting prompts---can yield human-equivalent performance in across a complex set of socio-historical codes. Than et al. \citeyear{than2025updating} similarly show that generative LLMs can replicate and augment traditional hand-coding workflows, particularly when they are treated as natural-language interlocutors within an iterative coding process. At the same time, method papers and practice guides now offer concrete recommendations for prompt structure, model selection, and validation protocols, positioning LLMs as serious candidates for large-scale annotation in computational social science and beyond \cite[e.g.,][]{tornberg2024best,ziems2024can}.

Yet early successes in LLM-assisted coding may obscure unresolved problems with reliability. 
Good test-set performance does not guarantee reliable labels when models are deployed at scale or on new corpora. Törnberg \citeyearpar{tornberg2024best} characterizes current LLM-based annotation practice as a methodological ``wild west,'' emphasizing that even small changes in prompts or data can produce unstable outputs. Ashwin et al. \citeyearpar{ashwin2023using} show that LLM annotation errors may not be random noise: they can be systematically correlated with respondent characteristics, inducing biased downstream inferences. Ziems et al. \citeyearpar{ziems2024can} likewise find that overconfident models can corrupt label sets in ways that standard ensembling does not fix and explicitly argue for human-in-the-loop controls rather than blind automation. Consensus has rapidly formed among methodologists that LLMs can advance social science only so long as they are tightly constrained by human-derived objectives and monitoring: what Sanaei and Rajabzadeh \citeyearpar{sanaei2025depth} have termed ``bounded-autonomy.'' Accordingly, existing guides  \cite[e.g.,][]{%dunivin2025scaling,
,than2025updating} focus on how to design and validate first-pass annotators of human-led research questions. Far less attention has been given to what researchers should do when a responsibly calibrated automated annotator---whether formally ``good enough'' or still below a desired threshold---produces too many structured errors for substantive use. This paper addresses that gap by treating post-hoc error analysis and subsequent error correction as their own stage in the annotation pipeline. We demonstrate how a secondary LLM critique step that leverages self-reflection to re-examine errors from a first-pass annotator and target recurring failure modes, substantially improving the effective quality of labels.

We characterize this secondary LLM critic as a narrow form of self-reflection. Instead of treating the first-pass code assignment as final, we ask a second model call to re-examine the original passage and the first model’s rationale, and then either confirm or overturn the label according to a small set of known failure modes. The model reflects on a prior model decision: it inspects how that decision was justified and checks whether it satisfies stricter criteria. Whereas self-reflection typically refers to an open-ended process of iterative refinement, we put a human in the loop: the model evaluates the initial annotation against researcher-defined semantic constraints informed by empirical error analysis and research goals. Note also that we distinguish LLM self-reflection from the broader hermeneutic reflexivity of the researcher; here, ``reflection'' refers to a model task in a critique stage of the annotation pipeline.

Self-reflection is one of several promising pathways to understanding and improving artificial intelligence capabilities. ``Self-Refine'' \cite{madaan2023self} and related approaches show that models can improve their own outputs by generating feedback and then revising their responses in an iterative feedback-refine loop. ``Reflexion'' \cite{shinn2023reflexion} and subsequent multi-agent systems extend this idea to agentic settings, where language agents verbally reflect on failures, store those reflections in memory, and use them to make better decisions in future episodes. Renze and Güven \citeyearpar{renze2024self} demonstrate that instructing LLM agents to reflect on incorrect answers and generate guidance for themselves significantly improves multiple-choice problem-solving performance. Other work pushes self-reflection inward: Kirchhof et al. \citeyearpar{kirchhof2025self} ask whether models can express their internal answer distributions in natural language, while Anthropic’s introspection studies \cite{lindsey2025emergent} probe whether models can report aspects of their own internal states. Finally, safety-oriented work suggests self-reflection can approach normative standards of conduct by having models critique and filter their own outputs \cite{liu2024self}. Across these lines, self-reflection entails a model generating commentary on its own behavior and then using that commentary to refine answers, calibrate confidence, or enforce constraints.

We advance the literature on LLM-assisted qualitative coding by incorporating targeted self-reflection into qualitative coding, where ``correctness'' is determined by human-derived codebooks, small gold-standard datasets, and manual error analyses. In our case study, the first-pass annotator achieves respectable test-set F1 yet systematically produces intolerably many false positives when run on the full data. Rather than train a new model or abandon automation, we design a secondary prompt that directs a critic model to perform a targeted self-reflection. The LLM critic reads the text and the first model’s rationale, checks for two recurring error types (misinterpretation and meta-discussion), and vetoes the label when all rationales are erroneous. Critically, LLM self-reflection is guided by a human-derived prompt, informed by manual error analysis, that orients the model to correct specific errors identified by human researchers. Because this critic operates only on positives, it is compute-efficient; because it is keyed to researcher-defined error modes, it brings the model’s self-reflection in line with the intended meaning of each code. The result is a boundedly-autonomous self-reflection that substantially improves the effective quality of LLM-assisted annotations in our corpus.

This paper begins by giving an overview of the methodology, our case study, and data. We developed this approach while annotating email messages for six classes of justifications in an \textit{in situ} group decision-making scenario: project evaluation by members of the Apache Software Foundation (ASF). We proceed to describing a post-hoc analysis of the LLM-annotations that produces a two-part error taxonomy. Most importantly, we provide and analyze the components of a working prompt for secondary LLM critique. The Results section then demonstrates evidence that this prompting strategy greatly improves annotations for four deficient codes. Finally, the discussion explores the role of error analyses in LLM-assisted coding and explains how to design a precision-first pipeline with the express intention of relying on LLM-critique for maximal performance.

\begin{figure}[t]
\centering
\resizebox{\columnwidth}{!}{%
\begin{tikzpicture}[
  node distance=10mm and 13mm,
  box/.style={draw, rounded corners, align=center, inner sep=3.5pt, minimum width=36mm},
  data/.style={draw, cylinder, shape border rotate=90, aspect=0.25, align=center, inner sep=4pt, minimum width=36mm},
  callout/.style={draw, dashed, rounded corners, align=left, inner sep=4pt, text width=35mm},
  arr/.style={-Latex, line width=0.6pt},
  darr/.style={-Latex, dashed, line width=0.6pt}
]

% Spine (vertical)
\node[data] (corpus) {\textit{ASF emails}\\\textbf{N=3,149}};
\node[box, below=of corpus] (s1) {\textbf{Stage-one: Annotator}\\Label + rationale};
\node[box, below=of s1] (pos) {S1 predicted positives};%\\(\,$\approx$15\%\,)};
\node[box, below=of pos] (s2) {\textbf{Stage-two: Critic}\\S1 labels + rationales\\Veto / confirm};
\node[box, below=of s2] (out) {\textbf{Final labels}\\(S2 outputs)};

\draw[arr] (corpus) -- (s1);
\draw[arr] (s1) -- (pos);
\draw[arr] (pos) -- (s2);
\draw[arr] (s2) -- (out);

% Evaluation callouts (right side, aligned)
\node[callout, right=15mm of s1] (gs1) {\textbf{Eval Phase 1}\\\textit{Enriched gold standard}\\\textbf{N = 120}\\Prompt development\\S1 validation};
\draw[darr] (gs1.west) -- (s1.east);

\node[callout, right=15mm of pos, yshift=-12mm] (audit) {\textbf{Eval Phase 2}\\\textit{S1 predicted positives}\\\textbf{N = 60 per code = 360}\\Error taxonomy\\
Prompt development \\ S2 validation};
\draw[darr] (pos.east) -- ++(6mm,0) |- (audit.west);
\draw[darr] (audit.west) -- ++(-6mm,0) |- (s2.east);    % informs/validates S2

\node[callout, right=15mm of out] (gs2) {\textbf{Eval Phase 3}\\\textit{Natural gold standard}\\\textbf{N = 150}\\ Compare S1 to S2};
\draw[darr] (gs2.west) -- (out.east);

\end{tikzpicture}%
}
\caption{Two-stage annotation pipeline with three evaluation phases. While the pipeline summarizes model execution on the corpus, the evaluation phases document the human-guided process of prompt development, error analysis, and validation that makes the workflow reliable in practice.}
\label{fig:pipeline_overview}
\end{figure}
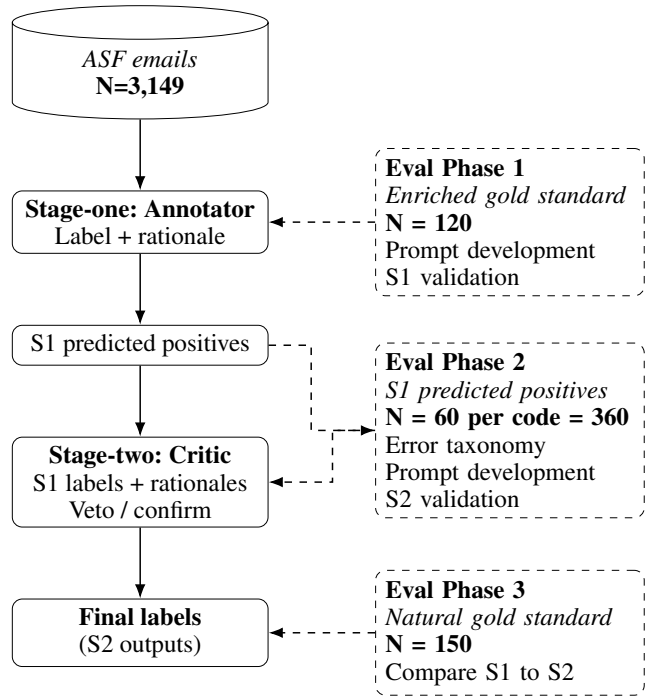

\def\arraystretch{1.4}%  1 is the default, change whatever you need
\begin{table*}[ht]
\centering
\begin{tabular}{ll}
\toprule

\multicolumn{1}{c}{\bf Code} & \multicolumn{1}{c}{\bf Description}\\
\midrule
Community Vitality & Health and activity of the project's developer and user community\\
Corporate Involvement & For-profit company support for a project\\
Cultural Alignment & Alignment with ASF's collaborative and open-source principles\\
Policy Compliance & Adherence to Apache policies and procedures.\\ 
Mentor Engagement &  Impact and involvement of project mentors\\
Technical and Market & Project's technological significance and appeal to potential users or contributors\\
\bottomrule

\end{tabular}
\caption{Categories and descriptions for six original substantive codes.} % title of Table
\label{table:code_listing}
\end{table*}

\section{Methodology}
Our study proposes a two-stage annotation workflow. In stage-one, an LLM applies a per-code prompt derived from an LLM-adapted qualitative codebook, producing a binary label and a brief rationale for each passage. In stage-two, a separate LLM call critiques only the stage-one \emph{predicted positives}. The critic model re-reads the original passage alongside the stage-one rationale and either confirms the label or vetoes it according to a compact decision policy keyed to observed failure modes. Figure~\ref{fig:pipeline_overview} summarizes the full workflow and how it connects to our evaluation design.

We validate and evaluate this two-stage workflow across three evaluation phases. In Evaluation Phase 1 an enriched gold-standard set used to confirm that stage-one prompting yields reasonable agreement with expert coding under controlled conditions; detailed per-code metrics are reported in appendix Table~\ref{table:primary_performance}. Evaluation Phase 2 audits stage-one predicted positives sampled from the full corpus (60 per code) to (i) characterize recurrent errors and (ii) evaluate the critic in the same decision regime it faces in deployment: distinguishing valid from invalid items within the pool of stage-one positives. In Evaluation Phase 3 a natural gold-standard set used to assess end-to-end performance after secondary critique under empirical base rates; because some codes are rare in a purely random sample, we report prevalence-corrected estimates that augments the gold standard with audited predicted-positive cases.

Below we describe the data and codebook, stage-one prompting and corpus-wide annotation, the audit procedure and resulting error taxonomy, the critic prompt and its evaluation on audited positives, and finally the full pipeline evaluation design.

\subsection{Codebook development and first-stage coding}
We developed six qualitative codes that target how the senior members of the Apache Software Foundation evaluate projects success after a probationary period. The codebook derives from a thematic analysis of 16 semi-structured interviews with ASF members, surfacing the criteria used in collective decision-making. Following Dunivin \citeyearpar{dunivin2025scaling}, we adapted the human codebook for LLM use by (i) tightening definitions and scope, and (ii) adding boundary clauses and negative examples to reduce near-misses. Abbreviated code descriptions are given by Table~\ref{table:code_listing} with complete code descriptions in appendix Section~\ref{app:code_descriptions}.

Two expert coders independently labeled a gold-standard set of 120 emails drawn from high-content project evaluation discussions. Disagreements were adjudicated to consensus. %Table~\ref{table:primary_performance} reports Cohen's $\kappa$, precision, recall, F1, and detected positive rate for the gold standard and GPT-4o on this set. F1 ranged from 0.74--0.77 for Cultural Alignment and Policy Compliance; 0.80--0.83 for Community Vitality, Mentor Engagement, and Technical and Market; and 1.00 for Corporate Involvement on this sample.

We ran separate per-code prompts (3,149×6 API calls, with 3,149 messages and 6 codes). Each prompt asked the model to issue a label and a brief rationale, which we retained for use by the stage-two critic. Decoding employed the following settings: \texttt{gpt-4o-2024-08-06; temperature = 0; top-p = 1}.

\begin{figure*}
    \centering
    \includegraphics[width=\textwidth]{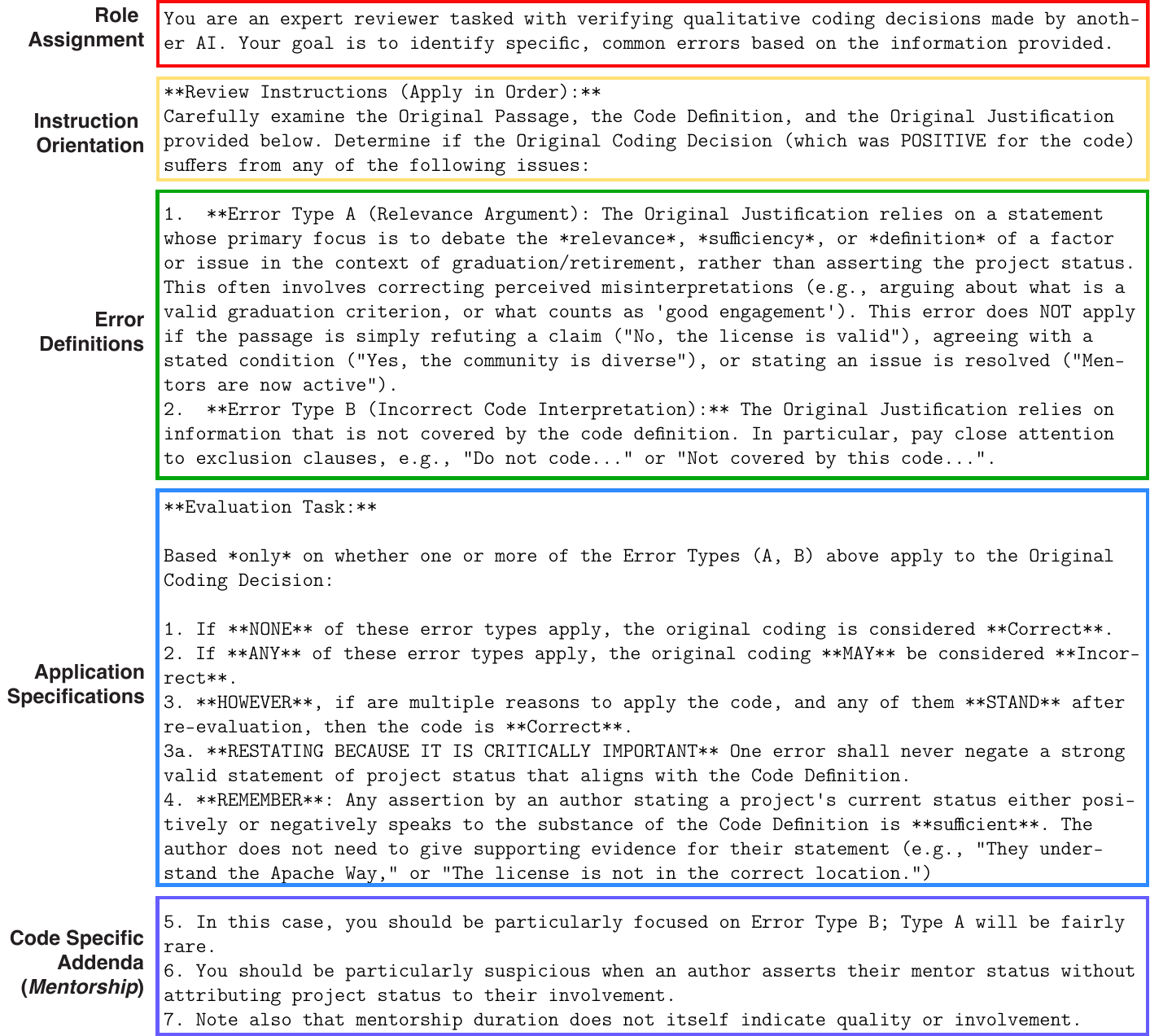}
    \caption{First and second layer of the prompt for secondary LLM critique for the Mentor Engagement code.}
    \label{fig:prompt_example_1}
\end{figure*}

\begin{figure*}
    \centering
    \includegraphics[width=\textwidth]{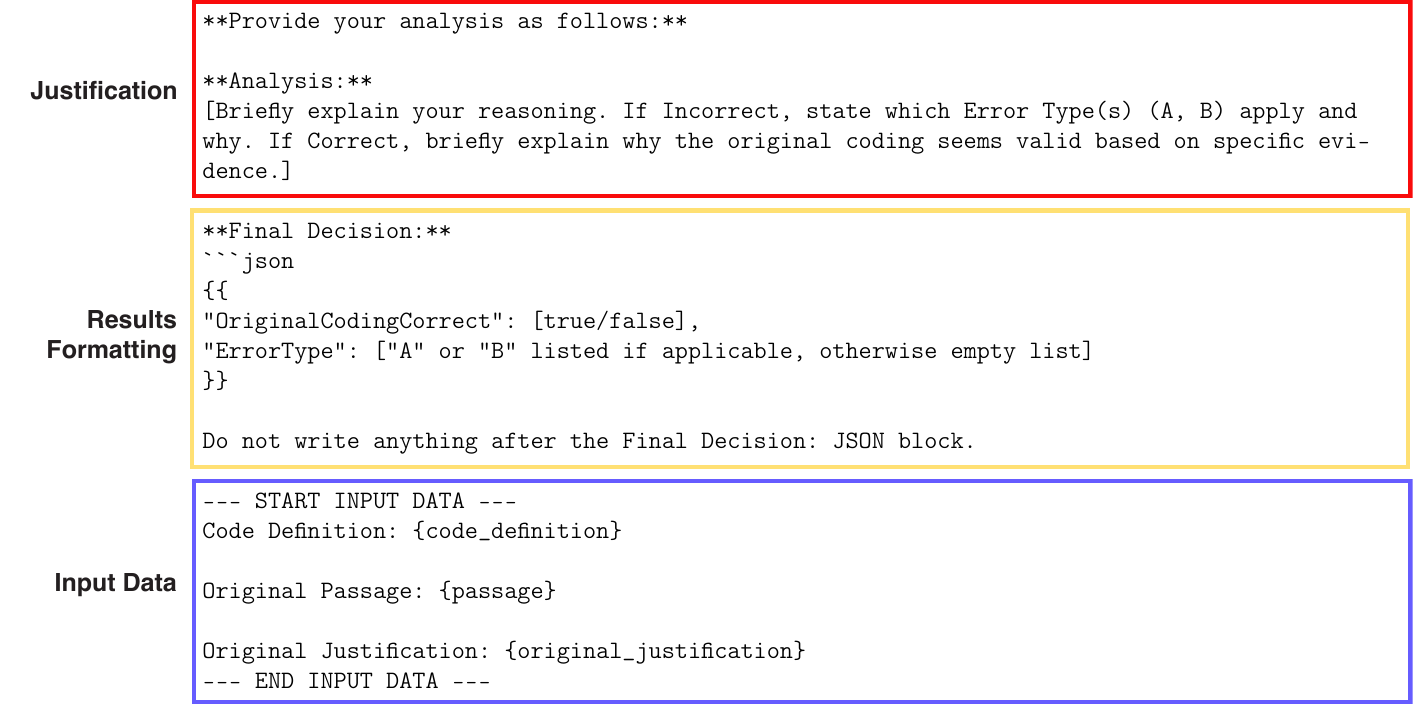}
    \caption{Third layer of the prompt for secondary LLM critique including the chain-of-thought justification, formatting instructions, and original passage and annotation for critique.}
    \label{fig:prompt_example_2}
\end{figure*}

\subsection{Error analysis: developing a classification schema for false positives}
LLM-assisted coding produced higher positive rates for several codes than predicted by our gold standard, underscoring the need for post-hoc quality checks even when test-set agreement is high. To diagnose where the first-pass model fails, one author audited a random sample of 60 stage-one positive passages per code drawn from the full pool of positives. For each passage, the rater rendered a human validity judgment (does the passage satisfy the code definition?) and assigned an error class when invalid.

During the error analysis, the human rater read the stage-one rationale. To preserve the integrity of the gold-standard validity labels, adjudication followed a basis-of-judgment rule: labels were determined solely from the passage content against the code definition; rationales were used only to locate candidate spans and to classify the error when invalid. This approach supports both aims---constructing a reference set and developing an error taxonomy---though we note a residual risk of anchoring toward the model’s framing \cite{schroeder2025just}.

%Across the six codes, we observed substantial false-positive rates among the reviewed positives for four codes (Table~\ref{table:error_analysis}): from 0.20 for Policy Compliance to 0.53 for Mentor Engagement (denominator = 60 per code). 
Multiple rounds of inductive coding yielded two recurrent error classes:

\begin{itemize}
    \item \textbf{Meta-discussion:} the passage is about a criterion (debating its relevance, sufficiency, or meaning) rather than using it as a justification for evaluation. This was especially common for Policy Compliance (e.g., debating whether a requirement is actually codified), where discussion of rules is frequent.
    \item \textbf{Misinterpretation:} the model violates the code definition (often ignoring explicit exclusions or boundary clauses; e.g., “Do not apply when a mentor’s perspective is merely mentioned unless it evaluates mentorship quality”).
\end{itemize}

%Table~\ref{table:error_analysis} reports, for each code, the counts and proportions of human-valid items and of each error class among the 60 reviewed positives for both the human adjudicator and the LLM-critic employing our ultimate prompting strategy detailed below.

\subsection{Prompting strategy for secondary LLM critique}
Our critic prompt has three layers. First, task framing states who the critic is and what they’re checking. Second, a decision policy gives the error taxonomy, a sufficiency rule, and common failure modes. Third, an input/output contract that specifies the inputs (code definition, passage, stage-one rationale) and the required analysis plus machine-readable output (JSON). The prompt’s structure and the phrasing are shown in Figures~\ref{fig:prompt_example_1}–\ref{fig:prompt_example_2}. Figure~\ref{fig:prompt_example_1} comprises the first and second layer gives the working definitions for each Error Type A (meta-discussion / “Relevance Argument”) and Error Type B (misinterpretation / “Incorrect Code Interpretation”). Figure~\ref{fig:prompt_example_2} gives the third and final layer including rationale request, output formatting, and the specific stage-one fields passed to the critic.

Critique prompting is more complex than first-line annotation. The critic must assess whether the cited reasoning actually supports applying the code definition to the passage, and the decision standard differs across stages. In stage-one the annotator labels positive if any clause (or combination of clauses) satisfies the definition. In stage-two, the critic overturns the positive only if no cited justification remains valid after re-evaluation; we call this ``the sufficiency rule.''

We initially used a decision-tree with rigid slots and highly structured output, but this brittleness apparently distracted the model from the global sufficiency check. We therefore simplified the constraints into a compact Application Specifications block (Figure~\ref{fig:prompt_example_1}), which states when to reject the original annotation and classify the error. Clause 3a explicitly restates the sufficiency rule to reduce over-rejection: one error does not negate a strong, valid statement of project status. We found that without this summary, the LLM was inclined to report errors even in cases where some part of the original justification was valid. Even with the clause the model did not always observe the sufficiency rule, and sometimes erroneously rejected the original coding. Note that this more sufficiency rule will be necessary in cases only in which a passage is likely to contain multiple plausible instances of a qualitative code; in cases where there is a maximum one instance, simpler logic can be used, reducing overall prompt complexity. Finally Clause 4 clarifies that the original speaker need not provide specific evidence in support of their justificatory logic (e.g., Cultural Alignment).

For some codes, these general instructions were insufficient to catch recurrent failure modes. In two of the four critiqued codes, we added brief Code-Specific Addenda (Figure~\ref{fig:prompt_example_1}), for example, reminding the critic that mere mention of mentor status does not satisfy Mentor Engagement unless it evaluates mentorship quality. These addenda mirror the earlier codebook-adaptation process and addressed systematic misinterpretations without increasing prompt complexity.

We used the same model parameters for stage-two as stage-one: \texttt{gpt-4o-2024-08-06; temperature = 0; top-p = 1}.

\def\arraystretch{1.4}%  1 is the default, change whatever you need
\begin{table*}[!ht]
\centering % used for centering table

\begin{tabular}{
  l
  *6{S[table-format=1.2, round-mode=places, round-precision=2]}
}
\toprule

& \multicolumn{3}{c}{\bf Human}
& \multicolumn{3}{c}{\bf LLM}
\\
\cmidrule(r){2-4}\cmidrule(l){5-7}
\multicolumn{1}{c}{\bf Code}
& \multicolumn{1}{c}{\bf MD Error}
& \multicolumn{1}{c}{\bf MI Error}
& \multicolumn{1}{c}{\bf Total}
& \multicolumn{1}{c}{\bf MD Error}
& \multicolumn{1}{c}{\bf MI Error}
& \multicolumn{1}{c}{\bf Total}
\\
\midrule
Community Vitality    & 0.05 & 0.08 & 0.12 & \multicolumn{1}{c}{---} & \multicolumn{1}{c}{---} & \multicolumn{1}{c}{---} \\
Corporate Involvement & 0.07 & 0.02 & 0.08 & \multicolumn{1}{c}{---} & \multicolumn{1}{c}{---} & \multicolumn{1}{c}{---} \\
Cultural Alignment    & 0.12 & 0.20 & 0.27 & 0.15 & 0.10 & 0.25 \\
Mentor Engagement     & 0.07 & 0.52 & 0.53 & 0.02 & 0.55 & 0.57 \\
Policy Compliance      & 0.17 & 0.10 & 0.20 & 0.08 & 0.03 & 0.12 \\
Technical and Market  & 0.03 & 0.45 & 0.47 & 0.00 & 0.38 & 0.38 \\
\bottomrule %inserts single line
\end{tabular}
\caption{Human and LLM critque of false positives after the first round of LLM-coding. N = 60 per code. Meta-discussion (MD): Debates the aspects of the theme, but does not speak to project status. Misinterpretation (MI): LLM misinterpreted the code description, i.e., the passage does not relate to the theme.
} % title of Table
\label{table:error_analysis} % is used to refer this table in the text
\end{table*}

\subsection{Full pipeline evaluation design}

To validate the full two-stage coding workflow under the conditions in which it is meant to be used, we evaluate performance on a new gold-standard set of 150 passages randomly sampled from the final corpus (Evaluation Phase 3). This evaluation serves two purposes. First, earlier development sets intentionally increased the density of positive cases to support prompt iteration, but such sampling does not reflect empirical base rates or qualitative content of the full corpus. Second, deployment-scale auditing revealed a recurring interpretive boundary case---\emph{meta-discussion}---in which participants debate the meaning, relevance, or sufficiency of an evaluation criterion rather than invoking it as a justification for project status. Because the analytic significance of these ``use vs.\ mention'' cases became clear only after observing systematic false positives at scale, the Phase 3 gold standard applies a tightened validity rule: a passage is coded positive only when the theme is used to support an evaluation-relevant claim about project status, not merely discussed in the abstract.

A practical complication is that several themes are rare enough that a purely random evaluation yields too few predicted positives to support stable estimates of how often model-identified positives are substantively valid. We therefore use a two-part design that preserves empirical base rates while improving the reliability of estimates for the subset of cases most consequential for downstream use (predicted positives). Concretely, we (i) use the random sample to estimate each theme's prevalence and each stage's overall positive rate, and (ii) supplement this with additional audited predicted-positive cases drawn from the model's positive set to better estimate the share of predicted positives that meet the tightened validity rule. This corresponds to a case--control style evaluation in which prevalence is estimated from a naturalistic sample while precision conditional on a positive prediction is estimated with greater power from the enriched predicted-positive set \cite{begg1983assessment,rothman2008modern}. We then combine these quantities to report performance measures that reflect expected behavior under empirical base rates. The reconstruction used to obtain prevalence-consistent confusion rates and derived metrics is provided in Appendix~\ref{app:prevalence}.

\section{Results}
We report results for a two-stage annotation pipeline in which a stage-one annotator labels each passage and a stage-two critic re-examines only stage-one positives. Accordingly, our results separate (i) \emph{stage-two} (self-reflective critic) performance on audited predicted positives, where the critic either confirms or vetoes a positive label, from (ii) \emph{full pipeline} performance after stage-two critique, evaluated under empirical prevalence.

Results are reported across three evaluation phases (see Figure~\ref{fig:pipeline_overview}). First, we validate stage-one prompting against an enriched gold standard to confirm that the LLM-adapted codebook yields reasonable agreement under controlled conditions; because these baseline validation results primarily establish prompt adequacy rather than the paper’s main contribution, detailed per-code metrics are reported in the Appendix (Table~\ref{table:primary_performance}). Second, we audit stage-one predicted positives to characterize recurrent error types (Table~\ref{table:error_analysis}) and to evaluate the critic on the same decision regime it faces in deployment (Table~\ref{table:error_correction}). Third, we evaluate end-to-end performance after secondary critique on an additional random gold standard (Table~\ref{table:final_comparison_consolidated}), using prevalence-corrected estimates that combine the random sample with audited predicted-positive cases.

\bgroup
\def\arraystretch{1.4}%  1 is the default, change whatever you need
\begin{table*}[!htbp]
\centering % used for centering table
\begin{tabular}{
  l
  *6{S[table-format=1.2, round-mode=places, round-precision=2]}
}
\toprule
& \multicolumn{2}{c}{\bf Detected FP Rate} & \multicolumn{4}{c}{}
\\
\cmidrule(r){2-3}
\multicolumn{1}{c}{\bf Code} & \multicolumn{1}{c}{\bf Human}
& \multicolumn{1}{c}{\bf GPT-4o}
& \multicolumn{1}{c}{\bf Cohen's $\kappa$}
& \multicolumn{1}{c}{\bf Precision}
& \multicolumn{1}{c}{\bf Recall}
& \multicolumn{1}{c}{\bf F1}
\\
\midrule
Community Vitality    & 0.12 & \multicolumn{1}{c}{---} & \multicolumn{1}{c}{---} & \multicolumn{1}{c}{---} & \multicolumn{1}{c}{---} & \multicolumn{1}{c}{---}\\
Corporate Involvement & 0.08 & \multicolumn{1}{c}{---} & \multicolumn{1}{c}{---} & \multicolumn{1}{c}{---} & \multicolumn{1}{c}{---} & \multicolumn{1}{c}{---}\\
Cultural Alignment    & 0.27 & 0.25 & 0.69 & 0.80 & 0.75 & 0.77\\
Mentor Engagement     & 0.53 & 0.57 & 0.60 & 0.79 & 0.84 & 0.82\\
Policy Compliance      & 0.20 & 0.12 & 0.69 & 1.00 & 0.58 & 0.74\\
Technical and Market  & 0.47 & 0.38 & 0.70 & 0.91 & 0.75 & 0.82\\
\bottomrule %inserts single line
\end{tabular}
\caption{Stage-two LLM critique validation (stage-one predicted positives)} % title of Table
\label{table:error_correction} % is used to refer this table in the text
\end{table*}
\egroup

\subsection{Error analysis of stage-one positives and stage-two prompt validation}
We begin with the evaluation phase that motivates and directly tests secondary critique: an audit of stage-one predicted positives drawn from the full corpus. For each code, we sampled 60 passages that stage-one labeled positive and adjudicated whether each passage truly satisfies the code definition. This audit estimates the false-positive rate \emph{among predicted positives} at deployment scale and provides a compact error taxonomy for prompt design.

Table~\ref{table:error_analysis} shows that stage-one over-produced positives for four codes, yielding substantial false-positive rates among predicted positives: 0.27 for Cultural Alignment, 0.20 for Policy Compliance, 0.53 for Mentor Engagement, and 0.47 for Technical and Market. In contrast, Community Vitality and Corporate Involvement had comparatively low false-positive rates in this audit (0.12 and 0.08), leaving limited headroom for improvement from positive-only critique. The audit also clarifies the dominant failure mechanisms. Misinterpretation errors (violations of the code definition) dominate for Mentor Engagement and Technical and Market, while meta-discussion (discussion \emph{about} a criterion rather than invocation \emph{as} a justification) is especially common for Policy Compliance. These two error modes account for the great majority of invalid positives observed in the audit and motivate the critic’s decision policy.

We next evaluate the secondary critic on the same decision regime it faces in deployment: distinguishing human-valid from human-invalid items \emph{within} the pool of stage-one positives. In this setting, the critic’s task is naturally framed as detecting false positives (human-invalid items), rather than re-classifying the full corpus. Table~\ref{table:error_correction} reports agreement with human adjudication and standard classification metrics for this critique-stage decision. Across the four audited codes targeted for correction, the critic achieves moderate agreement (Cohen’s $\kappa$ = 0.60--0.70; $F1$ = 0.74--0.82) while substantially reducing the number of invalid positives that would otherwise pass through unchallenged. Performance varies by code in ways that mirror the audit taxonomy: for Policy Compliance, the critic is highly conservative (precision = 1.00) but misses a portion of invalid positives (recall = 0.58), consistent with the difficulty of separating meta-discussion from criterion invocation; for Mentor Engagement, the critic identifies most invalid positives (recall = 0.84) but sometimes rejects borderline valid cases (precision = 0.79), reflecting the code’s heavy burden on correct interpretation of mentor-related claims.

\bgroup
\def\arraystretch{1.4}
\sisetup{
    round-mode = places,
    round-precision = 2,
    table-number-alignment = center
}

\begin{table*}[!ht]
\centering

\begin{tabular}{l *9{S[table-format=+1.2]}}
\toprule
& \multicolumn{3}{c}{\textbf{Detected Positive Rate}} & \multicolumn{3}{c}{\textbf{Cohen's $\bm{\kappa}$}} & \multicolumn{3}{c}{\textbf{F1}} \\
\cmidrule(lr){2-4} \cmidrule(lr){5-7} \cmidrule(lr){8-10}
{\bf Theme} & {\bf Gold} & {\bf S1} & {\bf S2} & {\bf S1} & {\bf S2} & {$\bm{\Delta}$} & {\bf S1} & {\bf S2} & {$\bm{\Delta}$} \\
\midrule
Community Vitality    & 0.2133 & 0.2467 & 0.2467 & 0.8953 & 0.8878 & -0.0075 & 0.9193 & 0.9128 & -0.0065 \\
Corporate Involvement & 0.0867 & 0.0933 & 0.0933 & 0.9343 & 0.9299 & -0.0044 & 0.9402 & 0.9362 & -0.0040 \\
Cultural Alignment    & 0.1133 & 0.1600 & 0.1067 & 0.8031 & 0.8687 & +0.0656 & 0.8293 & 0.8831 & +0.0538 \\
Mentor Engagement     & 0.0400 & 0.0467 & 0.0333 & 0.5261 & 0.7839 & +0.2578 & 0.5465 & 0.7918 & +0.2453 \\
Policy Compliance      & 0.2867 & 0.3267 & 0.3000 & 0.7562 & 0.8189 & +0.0627 & 0.8307 & 0.8720 & +0.0413 \\
Technical and Market  & 0.0467 & 0.0400 & 0.0333 & 0.4958 & 0.6821 & +0.1863 & 0.5175 & 0.6944 & +0.1769 \\
\bottomrule
\end{tabular}
\caption{Classification performance for stage-one (S1) and stage-two (S2) on 150 randomly sampled passages. F1 and Cohen's $\kappa$ are reconstructed under empirical prevalence using audits of stage-one-identified positives to stabilize precision estimates.}
\label{table:final_comparison_consolidated}
\end{table*}
\egroup
Taken together, this audit-phase evaluation shows (i) that stage-one deployment errors are concentrated in a subset of codes, (ii) that these errors follow a small number of recurrent mechanisms (misinterpretation and meta-discussion), and (iii) that a critic prompt keyed to these mechanisms can detect many invalid stage-one positives in the critique regime. We next test whether these critique-stage improvements translate into end-to-end gains under empirical prevalence on a prevalence-representative sample of the corpus.

\subsection{Full pipeline performance after secondary critique}
We next evaluate the end-to-end effect of secondary critique on a prevalence-representative sample of 150 randomly sampled passages from the final corpus. Because some themes are rare in a purely random sample, we report prevalence-corrected performance estimates that combine the random sample (to estimate empirical prevalence and detected-positive rates) with audited predicted-positive cases (to stabilize precision among positives); the reconstruction procedure is described in appendix Section~\ref{app:prevalence}.

Table~\ref{table:final_comparison_consolidated} shows that improvements concentrate in the codes for which stage-one over-inclusion was most damaging. Mentor Engagement improves from $\kappa=0.53$ to $0.78$ ($\Delta=+0.26$) and from $F1=0.55$ to $0.79$ ($\Delta=+0.25$). Technical and Market improves from $\kappa=0.50$ to $0.68$ ($\Delta=+0.19$) and from $F1=0.52$ to $0.69$ ($\Delta=+0.18$). More moderate gains occur for Cultural Alignment ($\Delta F1=+0.05$) and Policy Compliance ( $\Delta F1=+0.04$), consistent with the critic pruning false positives while occasionally over-rejecting borderline true positives. Community Vitality and Corporate Involvement are essentially unchanged, as they were not subject to stage-two (predicted positive rate remains the same); slight adjustments to performance result from prevalence correction.

Shifts in detected-positive rates mirror this pattern. For Cultural Alignment, secondary critique reduces the positive rate from 0.16 to 0.11, bringing within rounding error of the gold prevalence. Policy Compliance improves similarly (0.33$\rightarrow$0.30 vs.\ 0.29). For the rare themes, by contrast, detected-positive rates in a 150-item sample are necessarily noisy: a change of 0.0067 corresponds to a single passage, so the observed shifts (e.g., Mentor Engagement 0.047$\rightarrow$0.033; Technical and Market 0.040$\rightarrow$0.033) reflect changes of only one to two passages and should be read as directional rather than precisely estimated reductions in false positives. For these rare codes, the more reliable signal comes from the prevalence-corrected agreement metrics, which incorporate audited predicted-positive cases to stabilize precision estimates. Overall, across the four codes targeted for correction, secondary critique consistently prunes stage-one over-inclusion in a way that translates into higher deployment-oriented agreement and F1.

\section{Discussion}

Our results demonstrate the efficacy of a two-stage, precision-first pipeline that uses targeted self-reflection to improve the reliability of LLM-assisted qualitative coding. A first-pass annotator is tuned to be inclusive, paired with a secondary critic that revisits only the positives in light of an empirically derived error taxonomy and code definitions. In our case study, this design substantially reduced false positives for some of the most difficult codes, while maintaining acceptable agreement with human adjudication on a held-out sample. More generally, it shows how familiar practices of iterative codebook refinement, error analysis, and adjudication can be translated into a structured, scalable workflow in which LLMs participate in, rather than replace, qualitative interpretive labor.

We put forward this approach as a way to leverage automated self-reflection without granting the model open-ended autonomy over its own reasoning. Unlike many self-reflection or ``chain-of-thought'' methods, our critic is implemented as a separate, tightly constrained model call that is asked to sharpen an existing decision rather than to freely reflect on it. This has two advantages. First, it gives researchers explicit control over what the model reflects on: the critic prompt can be steered to address specific, empirically observed error modes in LLM interpretation, rather than relying on generic, hidden, or embedded instructions to ``think step by step.'' Second, because the critic operates only on a subset of cases and is asked for short, structured justifications, it is typically less computationally complex and produces shorter, more auditable outputs than more elaborate self-reflection or reasoning schemes. In this sense, our pipeline exemplifies a bounded-autonomy, human-in-the-loop orientation: self-reflection is present, but it is instrumented and directed by the research team.

At the same time, treating self-reflection in this targeted way foregrounds its hermeneutic character. The error taxonomy and critic instructions we develop are not merely technical fixes, but codified outcomes of iterative and post-hoc engagement with the corpus, in which researchers reconsider both their own coding rules and the model's readings of specific passages. Disagreements between annotator, critic, and human labels become occasions to renegotiate how codes should be understood in practice, and those negotiations are then operationalized in the critic prompt. In the remainder of the discussion, we consider strategies for pipelines that include critique coding, how our error taxonomy might generalize to other corpora and qualitative schemes, and the broader role of error analysis in LLM-assisted qualitative coding.

\subsection{Designing LLM-annotation strategy for compatibility with critique coding}

In many qualitative schemas, especially as codes multiply and definitions become more complex, true positives are sparse for any individual code (even if they are not rare at the passage-level). Under this prevalence regime, false positives are relatively dense among stage-one positives, while false negatives are sparse among stage-one negatives. Concentrating critique on positives therefore finds and fixes more errors per item and accelerates both error analysis and gold-standard adjudication \cite[cf.][]{he2009learning,saito2015precision}. A key limitation is that this efficiency advantage is conditional: when true positives are common or false negatives are non-trivial, a positive-only critic will systematically miss an important class of errors unless the workflow is extended to audit and critique negatives as well, which we address below in brief.

When positives are rare, stage-one should be configured for recall, relying on the stage-two critic to deliver precision. In practice, stage-one should over-include borderline positives; the critic then applies a sufficiency rule and overturns a positive only if no cited justification remains valid under the code definition. This cascaded, precision-at-accept posture parallels selective prediction / ``reject-option'' workflows in ML \cite{geifman2019selectivenet,franc2023optimal}. As reported in numerous early studies of LLM-assisted qualitative coding, prompt design can be empirically tweaked for more or less inclusive code interpretations \cite{white2023prompt,tornberg2024best}.

We can favor recall by (i) avoiding premature exclusions that suppress borderline cases, (ii) prompting for brief, text-tethered rationales so the critic can evaluate sufficiency (this also improves precision), and (iii) using reliable verbal or numerical confidence estimates from the model and lowering the threshold \cite{chirkova2025judge,gligoric2025can}. The pipeline remains precision-first because the critic coding is designed to filter false positives.

Positive-only critique also saves compute. In our case, the stage-one positive rate was 15\%, so the critic operated on 15\% of passages, an 85\% reduction in critique-stage API calls relative to re-scoring the full corpus. This compute advantage is not universal: it scales with the observed positive rate, so corpora or codes with higher base rates will see smaller savings.

Some corpora (or codes) have many true positives. Here, false negatives become non-trivial, and it is appropriate to conduct an error analysis of negative codings. A minimal procedure is:

\begin{enumerate}
    \item Sample a small set of stage-one negatives for the code.
    \item Estimate the FN rate on the sample; if it’s appreciable, prompt for negative critique.
    \item Run the critic with a flipped sufficiency rule: change negative→positive if any valid justification exists in the passage under the code definition (i.e., the stage-one rule, not the exclusion logic used to overturn positives).
\end{enumerate}
This can be done in conjunction with the process of false-positive pruning, or possibly combined into a single prompt, depending on the FP rate and error complexity.

\subsection{Developing an error taxonomy}
We observed two recurrent error modes in first-pass annotations: misinterpretation and meta-discussion. This schema is not exhaustive, but is likely to cover a large set of errors in other datasets. Misinterpretation covers violations of the code definition, particularly in cases where stated scoping conditions were not observed by the model, or definition drifts following re-evaluation. Meta-discussion captures a different failure: passages that are about a criterion (debating its relevance, sufficiency, or meaning) rather than using it as a justification in the evaluation at hand. These two modes map onto distinct mechanisms---definition/boundary drift versus use–mention confusions---and together explain every observed false positive in our corpus. Other corpora may surface additional dominant mechanisms (or different mixtures of mechanisms), so the taxonomy should be treated as a starting point that is re-derived through local audit rather than assumed to transfer wholesale.

In practice, a critic focused on misinterpretation often suffices to improve first-pass annotations. Because these errors reflect systematic frictions between the human codebook and the model’s reading, they are amenable to clarifying edits (tightened exclusions, boundary examples) and other code-specific addenda that the critic can apply consistently. Meta-discussion errors are more characteristic of dialogic or deliberative settings, where speakers frequently cite criteria only to question their applicability or to set rules of debate. In such contexts, a precision-first critic correctly rejects many of these cases; further improvement would likely follow from dialogic modeling (tracking how an utterance responds to prior turns) to distinguish mention from use. We see promise here, but the capabilities of current models and the scientific value of large-scale conversational analyses remain open questions.

Although these two modes have been broadly useful for us, other settings should surface different, more or less widely applicable errors. The practical stance is minimal-but-extensible: start with a small taxonomy that captures the dominant failure mechanisms, and introduce targeted subtypes only for recurrent misinterpretations within a specific code. Particularly slippery codes may warrant multiple narrowly defined subtypes rather than a single omnibus rule; attempting to catch every edge case at once tends to bloat prompts and degrade LLM reliability.

\subsection{The role of iterative and post-hoc error analyses in LLM-assisted qualitative analysis}

Error analysis in LLM-assisted coding extends familiar qualitative practices around training coders, revising codebooks, and adjudicating disagreement into a more explicitly dialogic, hermeneutic engagement among researchers, their data, and the model. Classic qualitative texts treat coder disagreement as an analytic resource: a prompt to surface ambiguities in the codebook, document interpretive decisions, and sometimes reconsider the research focus itself rather than simply forcing consensus or maximizing a reliability coefficient \cite[e.g.,][]{miles2020qualitative,oconnor2020intercoder,cofie2022eight}. Recent work on LLM-assisted qualitative research similarly frames models as partners in a three-way conversation with data, where model outputs invite renewed reading and theoretical reflection rather than supplying definitive answers \cite{dunivin2025scaling,hayes2025conversing}. From a hermeneutic perspective, prompts, model responses, and human judgments together instantiate a shifting ``hermeneutic contract'' between reader and text \cite{henrickson2025prompting}. In this light, examining discrepancies between human and model labels or rationales is not merely a matter of correcting model mistakes, but part of an ongoing process of renegotiating how coding rules, examples, and task instructions encode the meanings that researchers take to be at stake in the corpus.

It is useful to distinguish between \emph{iterative} error analysis during annotator strategy development and \emph{post-hoc} error analysis after a model has been deployed at scale. Iterative error analysis operates on relatively small development sets and plays a role analogous to training and calibrating a team of human coders: researchers inspect disagreements, refine prompts and task instructions, and, critically, decide whether discrepancies reflect a correctable misunderstanding of the existing schema or a deeper misalignment that calls for revising the schema itself (rather than merely adjusting the model to it). This is consistent with long-standing guidance that codebooks should be treated as living documents that evolve through cycles of application, discussion, and refinement, rather than as fixed instruments applied once and for all \cite{miles2020qualitative,saldana2021coding}. Recent work on LLM-assisted coding similarly emphasizes iterative adaptation of definitions and prompts so that the model’s operationalization of a code better approximates the interpretive practice of the research team, rather than treating the original codebook as sacrosanct and the model as a purely mechanical implementer  \cite[e.g.,][]{than2025updating,schroeder2025just}.

Post-hoc error analysis takes place after a first-pass annotator has been applied to a large corpus. At this stage, researchers can examine richer samples of model decisions, including borderline and contentious cases that are too rare to appear in small calibration sets. In our study, systematic post-hoc review of stage-one positives revealed that meta-discussion---instances where participants invoked a theme only to question its relevance, clarify procedural ground rules, or explicitly bracket it---constituted a substantial fraction of the model's false positives. These cases were sparsely represented in the development set and initially treated as edge cases, but they proved to be both common and structurally patterned at corpus scale. This prompted us to recognize that our original codebook and gold standard had not adequately foregrounded a key requirement: that a theme be invoked in service of making a substantive claim about project status. Post-hoc error analysis thus led not only to revising the model prompts and critic instructions, but also to retroactively tightening the human coding rules and correcting a subset of gold labels. More generally, this illustrates how error analysis at scale can expose systematic ambiguities in human coding practice that are difficult to detect in the relatively small and sparse samples typical of a test-set.

A final benefit of sustained error analysis is that it may reveal limits of the human ``gold standard'' as well as of the model. In long or dense passages, human coders are susceptible to fatigue and selective reading, sometimes missing segments that satisfy a code even when, in hindsight, the rationale for coding is clear. When disagreements between an LLM and the gold standard are examined carefully, some will turn out to reflect human oversight rather than model misinterpretation or inconsistency. In such cases, it may be appropriate to revise the gold labels in light of the model’s reading, treating the LLM as a consistent but fallible collaborator rather than a subordinate ``assistant'' whose output is only ever corrected downward. Taken together, iterative and post-hoc error analyses encourage a more symmetric treatment of disagreement, in which both human and model judgments are open to revision and the ultimate goal is not simply to optimize performance metrics, but to stabilize a coding practice that faithfully reflects the constructs under study.

\section{Conclusion} 
We have introduced a two-stage, precision-first pipeline in which a targeted self-reflective LLM-critic reduces false positives from an inclusive first-pass LLM-annotator, making nuanced qualitative codes more usable at scale. This design contributes to an emerging hermeneutic turn in computational text analysis, where machine intelligence encodes and operationalizes human interpretive frameworks and practices. As foundation models improve in their ability to read long, complex texts and provide better-calibrated confidence estimates, we expect the practical utility of multistage, self-reflective LLM-assisted text annotation to grow, enabling more precise and transparent qualitative coding while centering human interpretation and research goals.

\appendix

\iffalse
\section{Acknowledgments}
AAAI is especially grateful to Peter Patel Schneider for his work in implementing the original aaai.sty file, liberally using the ideas of other style hackers, including Barbara Beeton. We also acknowledge with thanks the work of George Ferguson for his guide to using the style and BibTeX files --- which has been incorporated into this document --- and Hans Guesgen, who provided several timely modifications, as well as the many others who have, from time to time, sent in suggestions on improvements to the AAAI style. We are especially grateful to Francisco Cruz, Marc Pujol-Gonzalez, and Mico Loretan for the improvements to the Bib\TeX{} and \LaTeX{} files made in 2020.

The preparation of the \LaTeX{} and Bib\TeX{} files that implement these instructions was supported by Schlumberger Palo Alto Research, AT\&T Bell Laboratories, Morgan Kaufmann Publishers, The Live Oak Press, LLC, and AAAI Press. Bibliography style changes were added by Sunil Issar. \verb+\+pubnote was added by J. Scott Penberthy. George Ferguson added support for printing the AAAI copyright slug. Additional changes to aaai2026.sty and aaai2026.bst have been made by Francisco Cruz and Marc Pujol-Gonzalez.
\fi
%\bibliography{aaai2026}
\bibliography{references_epjds}

@article{henrickson2025prompting,
  title={Prompting meaning: {A} hermeneutic approach to optimising prompt engineering with {ChatGPT}},
  author={Henrickson, Leah and Mero{\~n}o-Pe{\~n}uela, Albert},
  journal={AI \& SOCIETY},
  volume={40},
  number={2},
  pages={903--918},
  year={2025},
  publisher={Springer}
}

@article{shinn2023reflexion,
  title={Reflexion: {A}n autonomous agent with dynamic memory and self-reflection},
  author={Shinn, Noah and Labash, Beck and Gopinath, Ashwin},
  journal={arXiv preprint arXiv:2303.11366},
  year={2023}
}

@article{tornberg2024best,
  title={Best Practices for Text Annotation with Large Language Models},
  author={Törnberg, Petter},
  journal={arXiv preprint arXiv:2402.05129},
  year={2024}
}

@article{chirkova2025judge,
  title={LLM-as-a-Qualitative-Judge: Automated Interpretive Evaluation of Text Generation},
  author={Chirkova, Nadezhda and colleagues},
  journal={arXiv preprint arXiv:2506.09147},
  year={2025}
}

@article{he2009learning,
  title={Learning from imbalanced data},
  author={He, Haibo and Garcia, Edwardo A},
  journal={IEEE Transactions on Knowledge and Data Engineering},
  volume={21},
  number={9},
  pages={1263--1284},
  year={2009},
  publisher={Ieee}
}

@article{saito2015precision,
  title={The precision-recall plot is more informative than the {ROC} plot when evaluating binary classifiers on imbalanced datasets},
  author={Saito, Takaya and Rehmsmeier, Marc},
  journal={PloS one},
  volume={10},
  number={3},
  pages={e0118432},
  year={2015},
  publisher={Public Library of Science San Francisco, CA USA}
}

@inproceedings{geifman2019selectivenet,
  title={Selectivenet: {A} deep neural network with an integrated reject option},
  author={Geifman, Yonatan and El-Yaniv, Ran},
  booktitle={International Conference on Machine Learning},
  pages={2151--2159},
  year={2019}
}

@article{franc2023optimal,
  title={Optimal strategies for reject option classifiers},
  author={Franc, Vojtech and Prusa, Daniel and Voracek, Vaclav},
  journal={Journal of Machine Learning Research},
  volume={24},
  number={11},
  pages={1--49},
  year={2023}
}

@article{ziems2024can,
  title={Can large language models transform computational social science?},
  author={Ziems, Caleb and Held, William and Shaikh, Omar and Chen, Jiaao and Zhang, Zhehao and Yang, Diyi},
  journal={Computational Linguistics},
  volume={50},
  number={1},
  pages={237--291},
  year={2024},
  publisher={MIT Press One Broadway, 12th Floor, Cambridge, Massachusetts 02142, USA~…}
}

@article{than2025updating,
  title={Updating “{The Future of Coding}”: {Qualitative} coding with generative large language models},
  author={Than, Nga and Fan, Leanne and Law, Tina and Nelson, Laura K and McCall, Leslie},
  journal={Sociological Methods \& Research},
  volume={54},
  number={3},
  pages={849--888},
  year={2025},
  publisher={SAGE Publications Sage CA: Los Angeles, CA}
}

@article{sanaei2025depth,
  title={Depth and Autonomy: {A} Framework for Evaluating {LLM} Applications in Social Science Research},
  author={Sanaei, Ali and Rajabzadeh, Ali},
  journal={arXiv preprint arXiv:2510.25432},
  year={2025}
}

@article{madaan2023self,
  title={{Self-Refine}: {Iterative} refinement with self-feedback},
  author={Madaan, Aman and Tandon, Niket and Gupta, Prakhar and Hallinan, Skyler and Gao, Luyu and Wiegreffe, Sarah and Alon, Uri and Dziri, Nouha and Prabhumoye, Shrimai and Yang, Yiming and others},
  journal={Advances in Neural Information Processing Systems},
  volume={36},
  pages={46534--46594},
  year={2023}
}

@article{renze2024self,
  title={Self-reflection in {LLM} agents: Effects on problem-solving performance},
  author={Renze, Matthew and Guven, Erhan},
  journal={arXiv preprint arXiv:2405.06682},
  year={2024}
}

@inproceedings{kirchhof2025self,
  title={Self-reflective uncertainties: Do {LLMs} know their internal answer distribution?},
  author={Kirchhof, Michael and F{\"u}ger, Luca and Golinski, Adam and Dhekane, Eeshan Gunesh and Blaas, Arno and Williamson, Sinead},
  booktitle={ICML 2025 Workshop on Reliable and Responsible Foundation Models},
  year={2025}
}

@article{lindsey2025emergent,
  author={Lindsey, Jack},
  title={Emergent introspective awareness in large language models},
  journal={Transformer Circuits Thread},
  year={2025},
  url={https://transformer-circuits.pub/2025/introspection/index.html}
}

@article{liu2024self,
  title={Self-reflection makes large language models safer, less biased, and ideologically neutral},
  author={Liu, Fengyuan and AlDahoul, Nouar and Eady, Gregory and Zaki, Yasir and Rahwan, Talal},
  journal={arXiv preprint arXiv:2406.10400},
  year={2024}
}

@article{ashwin2023using,
  title={Using large language models for qualitative analysis can introduce serious bias},
  author={Ashwin, Julian and Chhabra, Aditya and Rao, Vijayendra},
  journal={Sociological Methods \& Research},
  volume={Online first},
  year={2025},
  publisher={SAGE Publications Sage CA: Los Angeles, CA}
}

@article{hayes2025conversing,
  title={“{Conversing}” with qualitative data: {E}nhancing qualitative research through large language models ({LLMs})},
  author={Hayes, Adam S},
  journal={International Journal of Qualitative Methods},
  volume={24},
  pages={16094069251322346},
  year={2025},
  publisher={SAGE Publications Sage CA: Los Angeles, CA}
}

@inproceedings{schroeder2025just,
  title={Just Put a Human in the Loop? {Investigating LLM}-Assisted Annotation for Subjective Tasks},
  author={Schroeder, Hope and Roy, Deb and Kabbara, Jad},
  booktitle={Findings of the Association for Computational Linguistics: ACL 2025},
  pages={25771--25795},
  year={2025}
}

@article{oconnor2020intercoder,
  title={Intercoder reliability in qualitative research: {Debates} and practical guidelines},
  author={O’Connor, Cliodhna and Joffe, Helene},
  journal={International Journal of Qualitative Methods},
  volume={19},
  pages={1609406919899220},
  year={2020},
  publisher={SAGE Publications Sage CA: Los Angeles, CA}
}

@book{miles2020qualitative,
  title     = {Qualitative Data Analysis: A Methods Sourcebook},
  author    = {Miles, Matthew B. and Huberman, A. Michael and Salda{\~n}a, Johnny},
  year      = {2020},
  edition   = {4th},
  publisher = {SAGE Publications},
  address   = {Thousand Oaks, CA},
  isbn      = {978-1506353074}
}

@article{cofie2022eight,
  title={Eight ways to get a grip on intercoder reliability using qualitative-based measures},
  author={Cofie, Nicholas and Braund, Heather and Dalgarno, Nancy},
  journal={Canadian Medical Education Journal},
  volume={13},
  number={2},
  pages={73--76},
  year={2022},
  publisher={{\'E}rudit}
}

@article{dunivin2025scaling,
  title={Scaling hermeneutics: {A} guide to qualitative coding with {LLMs} for reflexive content analysis},
  author={Dunivin, Zackary Okun},
  journal={EPJ Data Science},
  volume={14},
  number={1},
  pages={28},
  year={2025},
  publisher={Springer}
}

@book{saldana2021coding,
  title={The Coding Manual for Qualitative Researchers},
  author={Salda{\~n}a, Johnny},
  year={2021},
  edition = {4th},
  publisher={Sage Publishing},
  address={Thousand Oaks, California}
}

@inproceedings{gligoric2025can,
  title={Can unconfident {LLM} annotations be used for confident conclusions?},
  author={Gligori{\'c}, Kristina and Zrnic, Tijana and Lee, Cinoo and Candes, Emmanuel and Jurafsky, Dan},
  booktitle={Proceedings of the 2025 Conference of the Nations of the Americas Chapter of the Association for Computational Linguistics: Human Language Technologies (Volume 1: Long Papers)},
  pages={3514--3533},
  year={2025}
}

@article{white2023prompt,
  title={A prompt pattern catalog to enhance prompt engineering with Chat{GPT}},
  author={White, Jules and Fu, Quchen and Hays, Sam and Sandborn, Michael and Olea, Carlos and Gilbert, Henry and Elnashar, Ashraf and Spencer-Smith, Jesse and Schmidt, Douglas C},
  journal={arXiv preprint arXiv:2302.11382},
  year={2023}
}

@book{rothman2008modern,
  author    = {Rothman, Kenneth J. and Greenland, Sander and Lash, Timothy L.},
  title     = {Modern Epidemiology},
  publisher = {Wolters Kluwer Health},
  address   = {Philadelphia, PA},
  year      = {2008},
  edition   = {3rd},
  isbn      = {978-1451193282},
}

@article{begg1983assessment,
  title        = {Assessment of diagnostic tests when disease verification is subject to selection bias},
  author       = {Begg, C. B. and Greenes, R. A.},
  journal      = {Biometrics},
  volume       = {39},
  number       = {1},
  pages        = {207--215},
  year         = {1983},
  doi          = {10.2307/2530820},
}

\appendix % This automatically changes section numbering to A, B, C...

% Reset the table counter
\setcounter{table}{0}

% Prepend "A" to table numbering (e.g., Table A.1)
\renewcommand{\thetable}{A\arabic{table}}

% (Note: \appendix usually handles the main letter, this is for sub-level control)
\renewcommand{\thesection}{A\arabic{section}}

\setlength{\parindent}{0pt}
\section{Code Descriptions}\label{app:code_descriptions}

\subsection*{Community Vitality}

\textit{Definition:}
This code addresses the health and activity of the project's community. Consider any of the following factors sufficient:

    -- Recruitment and retention of contributors\\
    -- Frequency of releases\\
    -- Communication frequency and quality\\
    -- Passion and dedication of community members\\
    -- Openness to feedback and new contributions\\
    -- Substantial growth in committers or PPMC members

This code applies when a user is commenting about any of these factors generally. Mention of a specific release, committer invitation, or mailing-list thread is insufficient to apply this code.   

\subsection*{Corporate Involvement}

\textit{Full Title:} Company Support/Dominance

\textit{Definition:}
This code pertains to how company support can affect the financial aspects of a project's sustainability, particularly by contributing paid contributors. Consider the following factors:

-- The presence of corporate developers who support the project\\
-- Fears of domination or dependence on a single company\\
-- Financial impact on developer time and resources\\
-- Effect of financial situation on attracting or retaining contributors

Company support is a benefit to a project, but Apache is against development dominated by a single company.\\
Company support only relates to project development, not project adoption.

%\section{Stage-one annotator results}
\def\arraystretch{1.4}
\begin{table*}[h]
 % title of Table
\centering % used for centering table

\begin{tabular}{
  l
  *6{S[table-format=1.2, round-mode=places, round-precision=2]}
}
\toprule
& \multicolumn{2}{c}{\bf Detected Positive Rate} & \multicolumn{4}{c}{}\\
\cmidrule(r){2-3}
\multicolumn{1}{c}{\bf Code} & \multicolumn{1}{c}{\bf Gold Standard} & \multicolumn{1}{c}{\bf GPT-4o} & \multicolumn{1}{c}{\bf Cohen's $\bm{\kappa}$} & \multicolumn{1}{c}{\bf Precision} & \multicolumn{1}{c}{\bf Recall} & \multicolumn{1}{c}{\bf F1}
\\
\midrule
Community Vitality        & 0.39 & 0.39 & 0.72 & 0.83 & 0.83 & 0.83 \\
Corporate Involvement     &  0.05 &  0.05 & 1.00 & 1.00 & 1.00 & 1.00 \\
Cultural Alignment        &  0.075 & 0.09 & 0.67 & 0.69 & 0.78 & 0.74 \\
Mentor Engagement         &  0.066 &  0.075 & 0.81 & 0.78 & 0.88 & 0.83 \\
Policy Compliance          & 0.15 & 0.208 & 0.69 & 0.64 & 0.89 & 0.77 \\
Technical and Market      &  0.058 &  0.066 & 0.79 & 0.75 & 0.86 & 0.80 \\
\bottomrule %inserts single line

\end{tabular}
\caption{Stage-one LLM-adapted codebook validation (N = 120)}
\label{table:primary_performance} % is used to refer this table in the text
\end{table*}

\subsection*{Cultural Alignment}

\textit{Full Title:} Apache Cultural Alignment

\textit{Definition:}
This code refers to the project's alignment with Apache Software Foundation's collaborative and open-source principles. Attend to any policy documents in the prompt when responding. Consider the following factors:

    -- Consensus-driven approach vs. single-leader (BDFL) structure\\
    -- Open communication and public contributions\\
    -- Adaptation to Apache's community-driven model

The mention of presence or absence community engagement is not sufficient to count as Cultural Alignment. With regard to community engagement, it must refer to the accessibility to new contributors or the openness of communication.\\
Lack of project diversity due to single-company dominance does not count as Cultural Alignment. This is covered by the Company Support/Dominance code.\\
Do NOT apply when aspects of Apache's culture are merely reflected in the purpose or spirit of the vote under discussion such as a desire to reach consensus. Alignment must reflect the project's broader practices and culture outside of the present voting process.

\subsection*{Incubator Policy}

\textit{Full Title:} Incubator/Apache Policy Compliance

\textit{Definition:}
This code relates to the project's adherence to specific Apache Incubator policies and procedures. Attend to any policy documents in the prompt when responding. Consider the following factors:

-- Following Incubator rules and guidelines
-- Release procedures\\
-- Licensing and branding compliance\\
-- Proper commit procedures and account usage\\
-- Open communication about development decisions

Compliance or noncompliance with voting procedures are not covered by this code. 

\subsection*{Mentor Engagement}

\textit{Full Title:} Mentor Engagement/Influence

\textit{Definition:}
This code relates to the impact and involvement of project mentors. Consider the following factors:

-- Activity level of mentors\\
-- Guidance in adapting to Apache practices\\
-- Flexibility in applying Apache recommendations\\
-- Overall impact of mentorship on the project

Do NOT apply when a mentor's perspective is mentioned unless it describes the quality of mentorship in the project.\\
Do NOT assume anyone mentioned by name is a mentor unless it is explicitly stated that they are a mentor.\\
Do NOT apply when a mentor is mentioned merely in their capacity to facilitate voting on graduation. The email must broadly indicate the role of mentorship in the project.

\subsection*{Technical and Market}

\textit{Full Title:} Technical Relevance and Market Fit

\textit{Definition:}
This code pertains to the project's technological significance and its appeal to potential users or contributors. Consider the following factors:

-- Ability to solve real-world problems\\
-- Existence and size of target audience\\
-- Competitive position relative to existing solutions\\
-- Adoption rates or potential for adoption

It is not sufficient to apply this code when a user merely describes the purpose of the software; technical relevance applies only when the project is described as performing well in solving the problems it was designed to, or if it fails to solve these problems. ``Audience size,'' ``competitive position,'' and ``adoption rates'' or other similar criteria are not subject to this same stringent requirement as ``problem solving.''

\section{Prevalence-corrected performance calculation}\label{app:prevalence}

We compute prevalence-consistent performance metrics using a case--control style evaluation design in which prevalence is estimated from a naturalistic sample, while precision conditional on a positive prediction is estimated with greater power from an enriched set of predicted positives \cite{begg1983assessment,rothman2008modern}.

\paragraph{Estimated quantities.}
For each theme and set of labels $s$ (annotator S1 and critic S2), we estimate three probabilities:
prevalence $\hat{\pi}=P(y=1)$,
predicted-positive rate $\hat{r}_s=P(\hat{y}_s=1)$,
and positive predictive value $\widehat{\mathrm{PPV}}_s=P(y=1\mid \hat{y}_s=1)$.
We then reconstruct a \emph{normalized} confusion matrix for each stage, where $\widehat{\mathrm{TP}}_s,\widehat{\mathrm{FP}}_s,\widehat{\mathrm{FN}}_s,\widehat{\mathrm{TN}}_s$ are expected \emph{proportions} (summing to 1), not counts.

\paragraph{Reconstruction.}
For each stage $s$, we reconstruct the normalized confusion rates as:
\begin{align}
\widehat{\mathrm{TP}}_s &= \hat{r}_s\,\widehat{\mathrm{PPV}}_s, \\
\widehat{\mathrm{FP}}_s &= \hat{r}_s\,(1-\widehat{\mathrm{PPV}}_s), \\
\widehat{\mathrm{FN}}_s &= \hat{\pi} - \widehat{\mathrm{TP}}_s, \\
\widehat{\mathrm{TN}}_s &= 1 - \widehat{\mathrm{TP}}_s - \widehat{\mathrm{FP}}_s - \widehat{\mathrm{FN}}_s.
\end{align}
If desired, expected counts for a corpus of size $N$ are obtained by multiplying each term by $N$.

\paragraph{Derived metrics.}
From these reconstructed rates we compute prevalence-consistent recall $\widehat{\mathrm{Rec}}_s=\widehat{\mathrm{TP}}_s/\hat{\pi}$ and F1. Precision is $\widehat{\mathrm{Prec}}_s=\widehat{\mathrm{PPV}}_s$ by definition. Cohen's $\kappa$ is computed from the reconstructed agreement rate $\widehat{P_o}_s=\widehat{\mathrm{TP}}_s+\widehat{\mathrm{TN}}_s$ and the corresponding chance agreement
\begin{align}
\widehat{P_e}_s=\hat{r}_s\hat{\pi}+(1-\hat{r}_s)(1-\hat{\pi}),
\end{align}
so that $\widehat{\kappa}_s=(\widehat{P_o}_s-\widehat{P_e}_s)/(1-\widehat{P_e}_s)$.
All reported metrics should therefore be interpreted as estimates of expected deployment behavior under empirical prevalence, while benefiting from increased stability of $\widehat{\mathrm{PPV}}_s$ afforded by targeted audits of predicted positives.

\end{document}